\documentclass{llncs}
\usepackage{amsmath,amssymb}
\usepackage{graphicx}
\usepackage{enumerate}
\usepackage{verbatim}
\usepackage{color}
\usepackage{xcolor}
\usepackage{subfigure}
\usepackage{pstricks,pst-tree,pst-node}
\usepackage{listings}
\usepackage{tikz}
\usepackage{float}
                                     \newcommand{\rtoc}[2]{\mathrel{\smash{\overset{\raisebox{-2pt}{\scriptsize $#1$}}{\Rightarrow_c^{#2}}}}}

\newcommand{\cc}[1]{}

\begin{document}
\title{Generating Stack-based Access Control Policies}
\author{Xin Li\inst{1} \and Hua Vy Le Thanh\inst{2}}
\institute{The University of Tokyo, \\
\and University of Science - Ho Chi Minh City} \maketitle

\lstloadlanguages{
         Java
 }

\lstnewenvironment{code}[1][]%
  {\minipage{\linewidth}
   \lstset{basicstyle=\ttfamily\footnotesize,frame=single,#1}}
  {\endminipage}

\definecolor{dkgreen}{rgb}{0,0.6,0}
\definecolor{gray}{rgb}{0.5,0.5,0.5}
\definecolor{mauve}{rgb}{0.58,0,0.82}
\lstset{ %
  language=Java,                % the language of the code
  basicstyle= \footnotesize,           % the size of the fonts that are used for the code
  numbers=left,                   % where to put the line-numbers
  numberstyle=\tiny\color{gray},  % the style that is used for the line-numbers
  stepnumber=1,                   % the step between two line-numbers. If it's 1, each line
                                  % will be numbered
  numbersep=5pt,                  % how far the line-numbers are from the code
  backgroundcolor=\color{white},      % choose the background color. You must add \usepackage{color}
  showspaces=false,               % show spaces adding particular underscores
  showstringspaces=false,         % underline spaces within strings
  showtabs=false,                 % show tabs within strings adding particular underscores
  frame=single,                   % adds a frame around the code
  rulecolor=\color{black},        % if not set, the frame-color may be changed on line-breaks within not-black text (e.g. commens (green here))
  tabsize=2,                      % sets default tabsize to 2 spaces
  captionpos=b,                   % sets the caption-position to bottom
  breaklines=true,                % sets automatic line breaking
  breakatwhitespace=false,        % sets if automatic breaks should only happen at whitespace
  title=\lstname,                   % show the filename of files included with \lstinputlisting;
                                  % also try caption instead of title
  keywordstyle=\color{black}\textbf,          % keyword style
  commentstyle=\color{black},       % comment style
  stringstyle=\color{black}\textit,         % string literal style
  escapeinside={\%*}{*)},            % if you want to add a comment within your code
}

\begin{abstract}

The stack-based access control mechanism plays a fundamental role in the security architecture of Java and Microsoft CLR (common language runtime). It is enforced at runtime by inspecting methods in the current call stack for granted permissions before the program performs safety-critical operations. Although stack inspection is well studied, there is relatively little work on automated generation of access control policies, and most existing work on inferring security policies assume the permissions to be checked at stack inspection points are known beforehand. Practiced approaches to generating access control policies are still manually done by developers based on domain-specific knowledges and trial-and-error testing. In this paper, we present a systematic approach to automated generation of access control policies for Java programs that necessarily ensure the program to pass stack inspection. The techniques are abstract interpretation based context-sensitive static program analyses. Our analysis models the program by combining a context-sensitive call graph with a dependency graph. We are hereby able to precisely identify permission requirements at stack inspection points, which are usually ignored in previous study.
\end{abstract}

\section{Introduction}

Access control is often the first step to protect safety-critical systems. In modern Web platforms, such as Java-centric web applications or Microsoft .NET framework, applications comprise components from different origins with
diverse levels of trust. A \emph{stack-based access control}
mechanism is employed in an attempt to prevent untrusted codes
from accessing protected resources. Access control policies are
expressed in terms of \textit{permissions} (e.g., a permission can be ``writing the file C:/students\_grades.txt'') that are granted to codes grouped by different domains (e.g., www.jaist.ac.jp/faculty). Developers can set checkpoints through the Java API \textsf{CheckPermission(Permission)} in
their programs, and access control is
enforced dynamically at runtime by  \textit{stack inspection}. When stack inspection is triggered, the current call stack will be inspected in a top-down manner, and methods in the call stack are checked for granted permission\cc{ until a \textit{privileged} method is found}. A caller can be marked as \textit{privileged}, and stack inspection stops at such a caller. If all callers have the specified permission until a privileged method is found, access control is passed and stack inspection returns quietly, otherwise the program execution will be interrupted immediately.

\cc{From a practical perspective, such runtime inspection may cause a
considerable runtime overhead. If access control at some checkpoints always
succeed at runtime, the runtime overhead can be reduced by
removing them.}

\begin{example}[\textbf{Semantics of Java Stack Inspection}]
\label{ex}
Consider the code snippet in Fig. \ref{code} that we borrow from \cite{Geay2009} and modify it to make checkpoints of stack inspections explicitly called and to make the analysis scenario  more complicated.

\begin{figure}
\begin{lstlisting}[label={lst:sample_code},numbers=left,escapeinside={@}{@}]
public class Lib {
  private static final String dir = "C:";
  private static final String domain= "JAIST.AC.JP";
  public static void checkConnect(String host, int port) throws Exception {
    @\label{let:newPerm}@SocketPermission p1 = mkSocketPerm(host, port);
    @\label{lst:checkPerm1}@AccessController.checkPermission(p1);
    Priv op = new Priv(dir, logFile);
    @\label{lst:doPriv}@AccessController.doPrivileged(op);
  }
  public static Permission mkSocketPerm(String host, int port)  throws Exception {
    String hn = host +":"+port;
    @\label{lst:string3}@SocketPermission p2 = new SocketPermission(hn,"connect");
    return p2;   
  }
}
public class Priv implements PrivilegedExceptionAction {
  public Object run() throws Exception {
    String name = "/log.txt";
    @\label{lst:string0}@String fn = Lib.dir + File.separator + name.substring(1);
    checkAccess(fn);
  }
  public void checkAccess(String fn) throws Exception {
    FilePermission p3 = new FilePermission(fn,"write");
    @\label{lst:checkPerm2}@AccessController.checkPermission(p3);
  }
 }
public class Faculty {
  public void connectFaculty() throws Exception {
    @\label{lst:string1}@String host = Lib.domain.toLowerCase() + "/faculty";
    @\label{lst:socket0}@Socket s = Lib.checkConnect(host, 8080);
  }
}
public class Student {
  public void connectStudent() throws Exception {
    @\label{lst:string2}@String host = Lib.domain.toLowerCase() + "/student";
    @\label{lst:socket1}@Socket s = Lib.checkConnect(host, 8080);
  }
}
\end{lstlisting}
\label{code}\caption{An Example for Java Stack Inspection}
\end{figure}

There are two library classes $\textsf{Lib}$, $\textsf{Priv}$, and two application classes $\textsf{Faculty}$ and $\textsf{Student}$. At the beginning of program execution, Java VM assigns all classes hereby methods in them to a set of permissions specified by a security policy. At runtime, the two clients will require to connect to their corresponding domains by creating a socket (Line \ref{lst:socket0} and \ref{lst:socket1}, respectively). Such a request will trigger stack inspection at Line \ref{lst:checkPerm1} by the API \textsf{AccessController.checkPermission(Permission)} which takes a single parameter of type \textsf{Permission} or its subclasses. $\textsf{Student}$ is required to posses the permission \\
$~~~~~~~~~~perm_s: ``\textsf{SocketPermission(jaist.ac.jp/student:8080,connect)}''$\\
 and $\textsf{Faculty}$ is required to hold the permission \\
$~~~~~~~~~~perm_f: ``\textsf{SocketPermission(jaist.ac.jp/faculty:8080, connect)}''$.

Moreover, the socket construction process should be logged in \textsf{C:/log.txt} by the system for later observation.  A file access permission
$perm_a$:``\textsf{FilePermission (C:/log.txt, write)}''
is required on the system to perform this task, hereby another stack inspection is triggered at Line \ref{lst:checkPerm2}.
Note that $\textsf{Student}$ and $\textsf{Faculty}$ reside on the current call stack but should not posses $perm_a$. To avoid authorization failures while logging, \textsf{Lib} invokes the API \textsf{doPrivileged} (Line \ref{lst:doPriv}) from the class $\textsf{AccessController}$ with passing an instance $\textsf{op}$ of \textsf{Priv}, and by Java semantics,  \textsf{op.run()} will be executed with full permissions granted to its caller, and stack inspection stops at \textsf{checkConnect} without requiring $perm_a$ from clients of \textsf{Lib}.
\end{example}

Although stack inspection is well studied, there is relatively little work on automated generation of access control policies, and most existing work on inferring security policies assume the permissions to be checked at stack inspection points are known beforehand. Practiced approaches to generating access control policies are still manually done by developers based on domain-specific knowledges and trial-and-error testing. 
Since testing cannot cover all program runtime
behaviors, the application could malfunction due to accidental authorization failures given a
misconfigured policy. If a security policy  is too conservative, i.e., some codes are granted more permissions than necessary, it violates the  PLP (Principle of Least Privilege), and the codes become vulnerable points for malicious attacks.

To the best of our knowledge, the only existing analysis that attempted to automatically identify authorization requirements and generate access control policies for Java applications is \cite{Geay2009}. As shown in Example \ref{ex}, reasoning permissions demands  points-to analysis for identifying objects of \textsf{Permission} type, and string analysis for resolving string parameters of relevant security APIs.  As declaimed in their paper, they are the first to combine access rights analysis with string-analysis for deriving a precise security policy. The analysis consists in a context-sensitive library analysis and a context-insensitive library-client analysis, being tailored for effectively analysing production-level programs. It is not studied how to resemble analysis results seamlessly in access rights analysis.

In this work, we present a systematic approach to automated generation of access control policy for the given program that necessarily ensure it to pass stack inspection. The techniques are abstract interpretation based (whole-program) context-sensitive static program analysis. Our techniques consist in the following new features.
\begin{enumerate}[(i)]
\item By defining a shared abstract interpretation of program calling contexts, different context-sensitive analysis modules required in access rights analysis are glued in the same framework. The shared abstract program calling contexts also enable us to generate permissions involved in the program and identify permissions at stack inspection points, and hereby  to generate access control policies. 
\item Our program model is based on context-sensitive call graph rather than ordinary call graph. The analysis based on it handles dynamic features of Java languages like late binding more precisely. The program model is encoded as conditional weighted pushdown systems and the analysis algorithm is yielded by model checking. 
\item Our program model combines context-sensitive call graph with dependency graph that essentially encodes data flow of permission objects. The reason why call graph does not suffice is because permission objects can be created and referred to anywhere in the program, by either accessing the heap, i.e., field access, or by parameter passing of method calls that are finished before stack inspection. In either case, the data flow of permission objects is beyond the scope of the current call stack inspected by access control.
\end{enumerate}

The rest of the paper is organized as follows. Section \ref{preliminary} reviews conditional weighted pushdown systems. Section \ref{abstraction} defines an abstraction interpretation of program calling contexts. Section \ref{sec:permissions} presents permission generation. Section \ref{formalization} formalizes the problem of access control policy generation. Section \ref{realization} gives realization algorithms as model checking problems on conditional weighted pushdown systems. Section \ref{discussion} discusses how to lift existing analysis to fit the analysis framework. Section \ref{related} discusses related work, and Section \ref{conclusion} concludes the paper. 

\section{Conditional Weighted Pushdown Systems}
\label{preliminary}

A pushdown system is a variant of pushdown automata without input alphabet. \cc{, and thus is not used as language acceptors but a model of system computation behaviors.}

\begin{definition}
\label{pds}A \textbf{pushdown system} $\mathcal{P}$ is
$(P,\Gamma,\Delta, p_0, \gamma_0)$, where $P$ is a finite
set of control locations, $\Gamma$ is a finite stack alphabet,
$\Delta \subseteq P \times\Gamma \times P\times\Gamma^{*}$ is a
finite set of transitions, $p_0 \in P$ is the initial control location, and $\gamma_0 \in \Gamma$ is the initial stack content.
A transition $(p, \gamma, q, \omega) \in \Delta$ is written
as $\langle p,\gamma \rangle \hookrightarrow \langle q, \omega
\rangle$. A \textbf{configuration} is a pair
$\langle q, \omega \rangle$ with $q\in P$ and $\omega
\in\Gamma^{*}$. A set of configurations $C$ is \textbf{regular} if
$\{\omega ~|~ \langle p, \omega \rangle \in C\}$ is regular. A
relation $\Rightarrow$ on configurations is defined, such that
$\langle p, \gamma \omega' \rangle \Rightarrow \langle q,\omega
\omega' \rangle \text{for each}~\omega' \in \Gamma^{*}$ if
$\langle p, \gamma \rangle \hookrightarrow \langle q,\omega
\rangle$, and the reflective and transitive closure of $\Rightarrow$ is denoted by $\Rightarrow^*$.
\end{definition}

A pushdown system can be normalized by a pushdown
system for which $|\omega| \leq  2$ for each transition rule
$\langle p, \gamma \rangle \hookrightarrow \langle q, \omega
\rangle$~\cite{Schwoon2002}. We omit $p_0$ and $\gamma_0$ when they do not apply.

\begin{definition} \label{semiring}
A \textbf{bounded idempotent semiring} $\mathcal{S}$ is $(D,
\oplus,$ $ \otimes,$ $  \bar{0}, \bar{1})$, where $\bar{0}, \bar{1}
\in D$, and
\begin{enumerate}
\item $(D,\oplus)$ is a commutative monoid with $\bar{0}$ as its
unit element, and $\oplus$ is idempotent, i.e., $a \oplus a = a$
for all $a \in D$; \item $(D,\otimes)$ is a monoid with $\bar{1}$
as the unit element; \item $\otimes$ distributes over $\oplus$,
i.e., for all $a, b, c \in D$, we have \\
$a \otimes (b \oplus c) = (a \otimes b) \oplus (a \otimes c)$ and
$(b \oplus c) \otimes a = (b \otimes a) \oplus (c \otimes a)$;
\item for all $a \in D, a \otimes \bar{0} = \bar{0} \otimes a =
\bar{0}$; \item A partial ordering $\sqsubseteq$ is defined on $D$
such that $a \sqsubseteq b$ iff $a \oplus b = a$ for all $a, b \in
D$, , and there are no infinite descending chains in $D$.
\end{enumerate}
\end{definition}

By Def. \ref{semiring}, we have that $\bar{0}$ is the greatest
element. From the standpoint of abstract interpretation, PDSs
model the (recursive) control flows of the program, weight
elements encodes transfer functions, $\otimes$ corresponds to the reverse of
function composition, and $\oplus$ joins data flows. A weighted pushdown system (WPDS) \cite{Reps2005} is a generalized analysis framework for solving meet-over-all-path problems for which data domains comply with the bounded idempotent semiring.

\begin{definition}
A \textbf{weighted pushdown system} $\mathcal{W}$ is
$(\mathcal{P}, \mathcal{S}, f)$, where
$\mathcal{P}=(P,\Gamma,\Delta,  p_0, \gamma_0)$ is a pushdown
system, $\mathcal{S} = (D, \oplus, \otimes,$ $ \bar{0}, \bar{1})$ is a
bounded idempotent semiring, and $f: \Delta \rightarrow
D$ is a weight assignment function.
\end{definition}

Let $\sigma = [r_{0},...,r_{k}]$ with $r_{i} \in \Delta$ for $0
\leq i \leq k$ be a sequence of pushdown transition rules. A value
associated with $\sigma$ is defined by $val(\sigma) = f(r_{0})
\otimes ... \otimes f(r_{k})$. Given $c, c' \in P \times
\Gamma^*$, we denote by $path(c,c')$ the set of transition
sequences that transform configurations from $c$ into $c'$.

\begin{definition} \label{movp}
Given a weighted pushdown system $\mathcal{W} = (\mathcal{P},
\mathcal{S}, f)$ where $\mathcal{P}
=(P,\Gamma,\Delta,  p_0, \omega_0)$, and regular sets of
configurations $S, T \subseteq P \times \Gamma^{*}$, the
\textbf{meet-over-all-path} problem computes
$$\textsf{MOVP}(S, T) = \oplus \{ val(\sigma) ~|~ \sigma \in
{path}(s, t), s \in S, t \in T \}$$
\end{definition}

We refer $\textsf{MOVP}(S, T)$ by $\textsf{MOVP}(S,T,$ $\mathcal{W})$ when there are more than one WPDS in the context.
WDPSs are extended to \textit{Conditional WPDSs} in \cite{Li2010}, by
further associating each transition with regular languages that
specify conditions over the stack under which a transition can be applied.

\begin{definition} \label{cpds}
A \textbf{conditional pushdown system} is $\mathcal{P}_c$
= $(P, \Gamma, \Delta_c, $ $\mathcal{C},  p_0, \gamma_0)$, where
$P$ is a finite set of control locations, $\Gamma$ is a finite
stack alphabet, $\mathcal{C}$ is a finite set of regular languages
over $\Gamma$, $\Delta_c \subseteq P \times \Gamma \times
\mathcal{C} \times P \times \Gamma^{*}$ is a finite set of transitions, $p_0 \in P$ is the initial control location, and $\gamma_0 \in \Gamma$ is the initial stack content. A transition $(p, \gamma, L, q, \omega) \in \Delta_c$ is
written as $\langle p, \gamma \rangle \overset{L}{\hookrightarrow}
\langle q, \omega \rangle$.
A {relation} $\Rightarrow_c$ on configurations is
defined such that $\langle p, \gamma \omega' \rangle \Rightarrow_c
\langle q, \omega \omega' \rangle$ for all $\omega' \in
\Gamma^{*}$ if there exists a transition $r: \langle p, \gamma
\rangle \overset{L}{\hookrightarrow} \langle q, \omega \rangle$
and $\omega' \in L$ \cc{written as $\langle p, \gamma \omega' \rangle
\rtoc{}{} \langle q, \omega \omega' \rangle$}. The reflecxive and
transitive closure of $\Rightarrow_c$ is denoted by
$\Rightarrow_c^*$.\cc{ We define $cpre^*(C) = \{c' \mid c' \Rightarrow_c^* c, c \in C \}$ and $cpost^*(C) = \{c ' \mid c \Rightarrow_c^* c', c \in C\}$ for any $C \subseteq P\times \Gamma^*$.}
\end{definition}

\begin{definition} \label{cwpds}
A \textbf{conditional weighted pushdown system}
$\mathcal{W}_c$ is a triplet $(\mathcal{P}_c, \mathcal{S}, f)$, where
$\mathcal{P}_c = (P, \Gamma, \mathcal{C}, \Delta_c,  p_0, \gamma_0)$
is a conditional pushdown system, $\mathcal{S} = (D, \oplus,
\otimes, $ $\overline{0}, \overline{1})$ is a bounded idempotent semiring, and $f:
\Delta_c \rightarrow D$ is a weight assignment function.
\end{definition}

We lift the model checking problem on WPDSs in Definition \ref{movp} to
Conditional WPDSs by replacing the underlying system from WPDSs to Conditional WPDSs, and refer it by $\textsf{MOVP}$ as well.

\section{Abstract Interpretation of Calling Contexts}
\label{abstraction}

%\begin{definition}[\textbf{Program Points}]
We denote by $\mathcal{M}$ the set of methods in a program,
and by $\mathcal{L}$ the set of program line numbers. Let $CallSite \subseteq \mathcal{M} \times \mathcal{L}$  denote the set of call sites, such that $l$ contains a method
call for any $(m,l) \in CallSite$. In sequel, we will always use $\zeta$ to range over $CallSite$.

\begin{definition}[\textbf{Call Graph}]
A \textit{call graph} $G = (N, E, s)$ is a directed graph, where
$N \subseteq \mathcal{M}$ is the set of nodes, $E \subseteq
\mathcal{M} \times CallSite \times\mathcal{M}$ is the set of
edges, and $s \in N$ is the initial node with no incoming edges.
We write $n \rightarrow n'$ for $(n, \zeta, n') \in E$, and
$\rightarrow^*$ is the transitive and reflexive closure of
$\rightarrow$. In particular, $n_{check} \in N$ and $n_{priv}\in N$ denote the method \textsf{checkPermission} and
\textsf{doPrivileged} from the class
\textsf{AccessController}, respectively.
\end{definition}

\cc{A call graph $G$ is built by call graph construction algorithms which is known
to be cyclically dependent of points-to analysis. If call graph
construction detects that $m$ calls $m'$ at line $l$, we have
$(m,(m, l, c), m') \in E$.}

The calling contexts of a method $m$, thereby local variables
residing in the method, is the set of (possibly infinite) sequences of call sites leading to $m$ from the program entry.

\begin{definition}[\textbf{Calling Contexts}]
\label{context}
By $Context \subseteq CallSite^*$ we denote program calling contexts in terms of call site strings.
Given a call graph $G = (N,E,s)$, the calling contexts of a method
$m$ is defined by $\phi: \mathcal{M} \rightarrow 2^{Context}$:
\begin{multline*}
\phi(m) = \{ \zeta_{k}  \dots  \zeta_1 \zeta_0 \in Context \mid \exists k\in \mathbb{N}: m_0 = s,   m_{k+1} = m,  \\ (m_i, \zeta_i, m_{i+1}) \in E, \text{ for each } 0 \leq i \leq k\}
\end{multline*}
\end{definition}

Given a finite set $S = \{s_0, \dots, s_k\}$, we denote by $\Pi~S$ the set of permutations of $S$. For a word $\omega = s_{i_0} s_{i_1}
\dots s_{i_j} \in S^*$ where $0 \leq j \leq k$ and $0 \leq i_j \leq
k$, we define $\Sigma(\omega) = \{s_{i_0}, s_{i_1}, \dots,
s_{i_j}\}$ to be the set of symbols that appear in $\omega$.

\begin{definition}[\textbf{Abstract Calling Contexts}]
\label{abs_context}
By $AbsCtxt \subseteq 2^{CallSite}$ we denote the abstract program calling contexts as sets of call sites appearing along each call sequence.
\begin{itemize}
\item An abstraction function $\alpha: Context \rightarrow AbsCtxt$ on calling contexts is
defined by, for $c \in Context$, $\alpha(c) = \Sigma(c)$, and an abstraction function $\tilde{\alpha}: 2^{Context} \rightarrow 2^{AbsCtxt}$ on sets of calling contexts is defined by, for $c \subseteq Context$,
%\begin{multline*}
$$\tilde{\alpha}(c) =\{ \Sigma(ctxt) \mid ctxt \in c, \text{and } cs' \notin \tilde{\alpha}(c) \text{ if } cs' \subset cs \text{ and } cs \in \tilde{\alpha}(c)\}$$
%\end{multline*}
\item A concretization function $\gamma: AbsCtxt \rightarrow 2^{Context}$
is defined by $\gamma(C) = \bigcup_{C' \subseteq C} \Pi~C'$ for $C \in AbsCtxt$, and the powerset extension of $\gamma$ is denoted by
$\tilde{\gamma}: 2^{AbsCtxt} \rightarrow 2^{Context}$.
\end{itemize}

The abstract calling contexts of a method $m$ is defined by a mapping
$\phi_{meth}: \mathcal{M} \rightarrow 2^{AbsCtxt}$, such that $\phi_{meth}(m) = \tilde{\alpha}(\phi(m))$.
\end{definition}

Let $\leq$ be a binary relation over $Context$ such that $ctxt \leq ctxt'$ for any $ctxt, ctxt' \in Context$ if $\Sigma(ctxt) \subseteq \Sigma(ctxt')$. We define $c \leq c'$ for $c, c'\subseteq Context$, if for each $ctext \in c$, there exists $ctext' \in c'$ such that $ctext \leq text'$. It is not hard to see that, for $c \subseteq Context$, $cs \subseteq AbsCtxt$, $\tilde{\alpha}(c) \subseteq cs$ iff $c \leq \tilde{\gamma}(cs)$, and we can hereby conclude with Theorem \ref{galois}.

\begin{theorem}
\label{galois}
$(2^{Context}, \tilde{\alpha}, \tilde{\gamma}, 2^{AbsCtxt})$ is a Galois connection. \qed
\end{theorem}

\begin{example}
Given a call graph $G=(N,E,s)$ where $N = \{m_1, \dots, m_4\}$ and $E = \{(m_i, \zeta_i, m_{i+1}) \mid 1 \leq i \leq 3\} \cup \{(m_3, \zeta_4, m_2), (m_2, \zeta_5, m_4)\}$\cc{ with $\zeta_i = (m_i, l_i, c)$ for $1 \leq i \leq 3$, $\zeta_4 = (m_3, l_4)$ and $\zeta_5 = (m_2, l_5, c)$}. We have $\phi(m_4) = \{\zeta_3(\zeta_2 \zeta_4)^* \zeta_2\zeta_1, \zeta_2 \zeta_1\}$, and $\phi_{meth}(m_4) = \{\{\zeta_i \mid i = 1, \dots, 4\}, \{\zeta_1, \zeta_2\}\}$. As shown in this example, the design of $\tilde{\alpha}$ concerns the situation of recursive calls.
\end{example}

\section{The Analysis Framework}

\begin{figure}
\begin{center}
\includegraphics[scale=0.4]{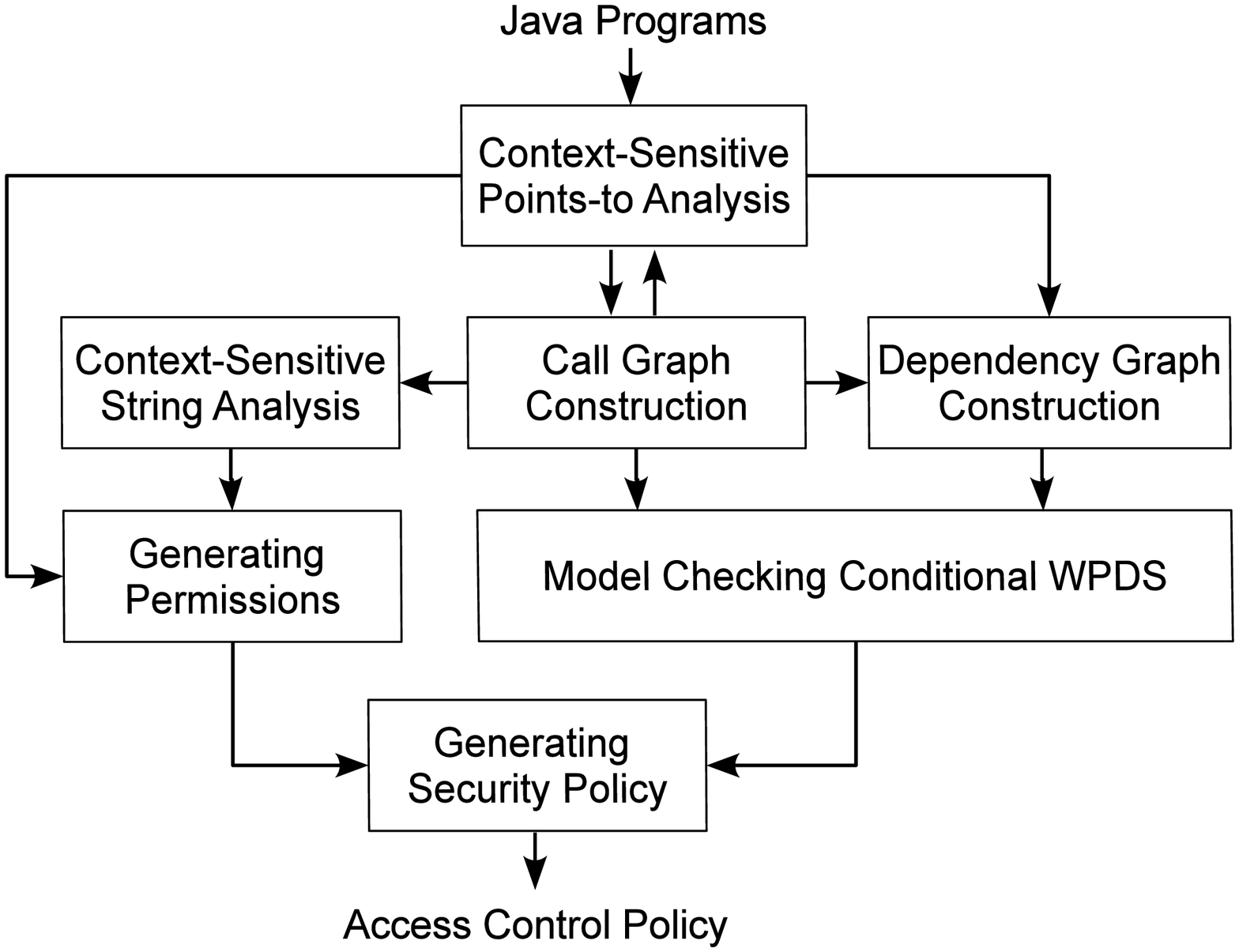}
\caption{The Overall Structure of the Analysis Framework}\label{structure}
\end{center}
\end{figure}

The overall structure of the analysis framework is shown in Figure \ref{structure}.  It consists of analysis modules shown in rectangles. Context-sensitive string and points-to analysis, call graph construction and dependency graph construction are pre-assumed and defined in this section, and we discuss in Section \ref{discussion} on how to adapt the off-the-shelf algorithms to fit the analysis framework.
 The generation of permissions and access control policies is given in the next sections. All analysis modules in Figure \ref{structure} are glued by means of a shared abstraction interpretation of program calling contexts given in Section \ref{abstraction}. In the rest of the paper, we will use Example \ref{code} as a running example in all examples.

%\subsection{Pre-assumed Analysis}

\begin{definition}[\textbf{Context-Sensitive Points-to Analysis}]
Given a reference variable $v$ of a method $m$, a context-sensitive points-to analysis, denoted by $\texttt{pta}(v)$,
\begin{enumerate}[(i)]
\item returns the finite set of abstract heap objects that $v$ may refer to at runtime under certain calling contexts; and 
\item each object in $ \texttt{pta}(v)$ is represented as a triplet $({Type},loc,c)$, where $Type$ is its runtime type, $loc$ is its allocation site, and $c \in \phi_{method}(m)$ is the calling contexts under which the object is allocated\cc{, and (iii) $\bigcup_{(type,loc,c)\in pta(v)} \{c\} \subseteq \phi_{meth}(m)$ for any reference variable $v$ of $m$}.
\end{enumerate}
\end{definition}

\begin{definition}[\textbf{Context-Sensitive String Analysis}]
 Given a string variable $v$ of the method $m$, a context-sensitive string analysis, denoted by $\texttt{sa}(v)$, 
 \begin{enumerate}[(i)]
 \item returns the finite set of string constants that $v$ may contain at runtime under certain calling contexts; and 
 \item each element in $\texttt{sa}(v)$ is represented as a pair $(sv, c)$, where $sv$ is the string value and $c \in \phi_{meth}(m)$ is the calling contexts under which $sv$ is constructed\cc{, and (iii) $\bigcup_{(sv,c)\in sa(v)} \{c\} \subseteq \phi_{meth}(m)$ for any string variable $v$ of $m$}.
 \end{enumerate}
\end{definition}

\cc{There are extensive research on context-sensitive points-to analysis and (mostly context-insensitive) string analysis. We discuss on how to adapt the off-the-shelf analyzers to context-sensitive counterparts required by our analysis.}
\cc{
The two dominating approaches to obtaining context-sensitivity of program analysis are known as \emph{context-cloning} and \emph{context-stacking}. Context-cloning resembles to inline expansion \cc{that copies the called procedures at each call site if possible} and as such has an inherent limit in handling recursive procedural calls. By context-stacking, we mean to model the program as a pushdown system and the analysis problem as model checking problems, e.g., in the framework of WPDSs in our setting. Since the stack of pushdown systems are unbounded, it naturally models recursive procedure calls.}

%\subsection{Program Model}

\begin{definition}[\textbf{Context-Sensitive Call Graph}]
\label{cs_callgraph}
A context-sensitive call graph $G_{cs} = (G, \phi_{edge})$
consists of a call graph $G = (N,E,s)$ and a mapping $\phi_{edge}:
E \rightarrow 2^{AbsCtxt}$, such that for each node $n \in N$,
\begin{itemize}
\item $\phi_{edge}(e) \subseteq \phi_{meth}(n)$ for each edge $e
= (n, \zeta, n') \in E$; \item $\bigcup_{e = (n, \zeta, n') \in E} \phi_{edge}(e) = \phi_{meth}(n)$.
\end{itemize}
We define a mapping $\phi_{route}: (\rightarrow^*) \rightarrow
2^{AbsCtxt}$ by, for each $n \rightarrow^{i} n'$,
\[
\phi_{route}(n
\rightarrow^i n') =
\left\{%
\begin{array}{ll}
\phi_{edge}(n \rightarrow n') & \hbox{if } i = 1 \\
%\\
\{c \cup c' \mid \exists n'' \in N: c \in \phi_{route}(n \rightarrow^{i-1}
n''), & \\
\quad \quad \quad c' \in \phi_{edge}(n'' \rightarrow n')\} & \hbox{if } i > 1  \\
\end{array}%
\right.
\]
\end{definition}

In Java, due to dynamic dispatch, the target method of a dynamic dispatch depends on the runtime type of receiver objects. A precise call graph construction computes virtual call targets separately for each calling contexts, and yields a context-sensitive call graph. The calling contexts under which a call edge and a call path is feasible are characterised by $\phi_{edge}$ and $\phi_{route}$ in Def. \ref{cs_callgraph}, respectively.  An algorithm for possibly constructing $G_{cs}$ is given in \cite{Li2010}.

\begin{definition}[\textbf{Dependency Graph}]
Given a program in SSA (Static Single Assignment) form. Let $T_{perm}$ denote the class (or type) \texttt{Permission} or any
of its subclasses. Let $\mathcal{L}_{alloc} \subseteq
\mathcal{L}$ be the set of program lines that allocate objects of
$T_{perm}$, and let $AllocPerm \subseteq \mathcal{M} \times
\mathcal{L}_{alloc}$.

A dependency graph $G_{dep}$ of the program is a directed graph $
(N_{dep}, E_{dep}, S_{dep})$, where $N_{dep} \subseteq
\mathcal{M}\times \mathcal{L}$ is the set of nodes, $E_{dep}
\subseteq N_{dep} \times N_{dep}$ is the set of edges, $S_{dep} =
AllocPerm$ is the set of initial nodes with no incoming edges.
Moreover, $E_{dep}$ is the smallest set that contains $(n, n')$
where $n = (m,l)$ and $n'=(m',l')$ if the variable of type
$T_{perm}$ defined in $l$ is used in $l'$. 

We further denote by $E_{inter} \subseteq E_{dep}$ the edges encoding (i) either statements of method invocation that pass arguments of type $T_{perm}$ or (ii) return statements that return values of type $T_{perm}$. 
\end{definition}

\begin{figure}[h]
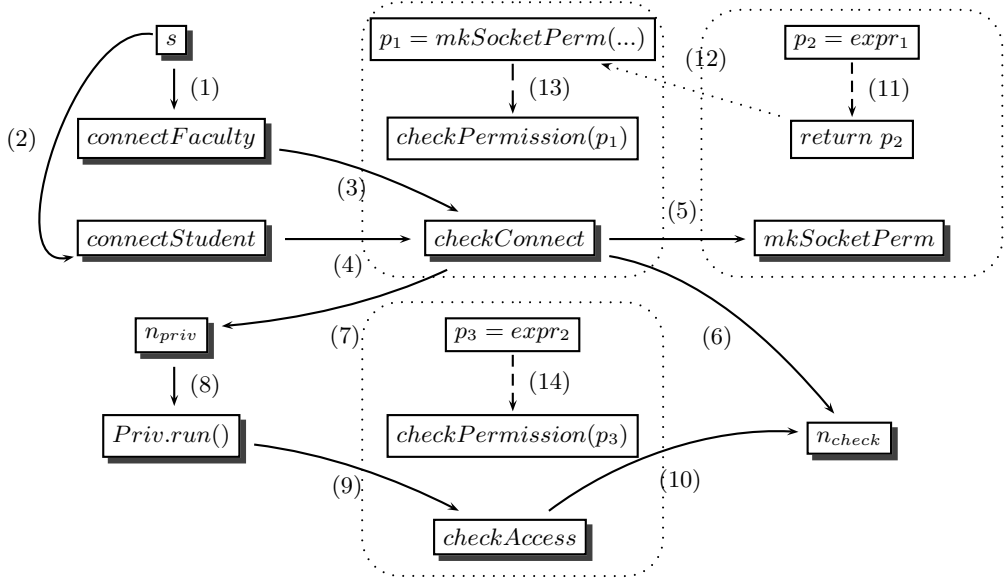

\centering
\[
%\psset{arrows=->,nodesep=10pt}
\begin{psmatrix}[rowsep=0.7cm,colsep=1.3cm]
\psshadowbox{s} &  \psframebox{p_1 = mkSocketPerm(...)}&  \psframebox{p_2 = expr_1}\\
\psshadowbox{connectFaculty} & \psframebox{{checkPermission}(p_1)}&\psframebox{return~ p_2}\\
\psshadowbox{connectStudent}  &  \psshadowbox{checkConnect} & \psshadowbox{mkSocketPerm}\\
\psshadowbox{n_{priv}} & \psframebox{p_3 = expr_2} & \\
\psshadowbox{Priv.run()}&  \psframebox{checkPermission(p_3)} & \psshadowbox{n_{check}}\\
& \psshadowbox{checkAccess} & \\
\psset{arrows=->,nodesep=3pt}
\ncline{1,1}{2,1}\naput{(1)}
\ncarc[arcangle=-100,nodesep=1pt]{1,1}{3,1}\tlput{(2)}
\ncarc[arcangle=20]{3,2}{5,3}\tlput{(6)}
\ncarc[arcangle=20]{6,2}{5,3}\nbput{(10)}
\ncarc[arcangle=10]{2,1}{3,2}\tlput{(3)}
\ncline[nodesep=5pt]{3,1}{3,2}\nbput{(4)}
\ncline{4,1}{5,1}\naput{(8)}
\ncarc[arcangle=10]{3,2}{4,1}\naput{(7)}
\ncline{3,2}{3,3}\taput{(5)}
\ncarc[arcangle=10]{5,1}{6,2}\tlput{(9)}
\ncbox[linearc=.5,boxsize=2,linestyle=dotted,nodesep=.2]{1,2}{3,2} %
\ncbox[linearc=.5,boxsize=2,linestyle=dotted,nodesep=.2]{4,2}{6,2} %
\ncbox[linearc=.5,boxsize=2,linestyle=dotted,nodesep=.2]{1,3}{3,3}%
\psset{arrows=->,nodesep=1pt,linestyle=dashed} %
\ncline{1,3}{2,3}\naput{(11)}
\ncline{1,2}{2,2}\naput{(13)}
\ncline[linestyle=dotted]{2,3}{1,2}\nbput{(12)}
\ncline{4,2}{5,2}\naput{(14)}
\end{psmatrix}
\]
\caption{The Call Graph and Dependency Graph for Example 1}
\label{graph}
\end{figure}

\begin{example}
The dependency graph and call graph of Example 1 is given in Figure \ref{graph}, where $expr_1$ abbreviates ``new SocketPermission(hn,``connect")", and $expr_2$ abbreviates ``new FilePermission(fn, ``write'')''. Rectangles and dashed lines represent nodes and edges of dependency graph, respectively, and dotted lines represent edges from $E_{inter}$ specifically. Rectangles with shadow and solid lines represent nodes and edges of call graph, respectively. Dotted lines are used to group nodes of call graph and dependency graph if they correspond to the same method. Edges of both dependency graph and call graph are labelled with numbers. We also label call sites following the labels of edges, such that an edge $(i)$ refers to $(m, \zeta_{(i)}, m')$ for some $m$ and $m'$.

We show how to generate a context-sensitive call graph, i.e., $\phi_{edge}$, given points-to analysis. Since $n_{check}$, $n_{priv}$, $checkConnect$, and $mkSocketPerm$  are static methods, these methods can be always called. The program entry point do not depend on previous calling contexts (that is empty) to dispatch a method invocation. \cc{Moreover, the same method checkAccess is shared by class ProPriv and Priv, and is called by Priv or ProPriv under no conditions though dynamically dispatched.} Therefore we have
$$\phi_{edge}((i)) = \emptyset \text{ for } i \in \{1,2,3,4,5,6,7,10\}$$ 
$Priv.run()$ is dynamically dispatched from line 8, depending on the runtime type of object $op$. Given $\texttt{pta}(op) = \{(Priv, 7, \{\zeta_{(1)}, \zeta_{(3)}\}),$ $ (Priv, 7, \{\zeta_{(2)}, \zeta_{(4)}\})\}$, we have 
$$\phi_{edge}((8)) = \{\{\zeta_{(1)}, \zeta_{(3)}\}, \{\zeta_{(2)}, \zeta_{(4)}\}\}$$
$$\phi_{edge}((9)) = \{\{\zeta_{(1)}, \zeta_{(3)}, \zeta_{(8)}\}, \{\zeta_{(2)}, \zeta_{(4)},\zeta_{(8)}\}\}$$
\end{example}

\section{Generating Permissions}
\label{sec:permissions}

All classes, hereby methods and
program points,  in a protection domain are granted the
same set of permissions. All methods belonging to the system domain, e.g., the
method \textsf{AccessController}.\textsf{doPrivileged}, are granted all permissions. Based on previously defined context-sensitive points-to and string analysis,  we show in this section how permissions that are possibly involved in the given program are generated. The result is an over-approximation of exact permissions appearing at runtime given sound string and points-to analysis. 

\begin{definition}[\textbf{Access Control Policy}]
Let $Domain$ denote the set of protection domains, and
$Perms$ denote the set of permissions. We denote by $\texttt{dom}: \mathcal{M} \rightarrow Domain$ the mapping from methods to their protection domains, and $\texttt{perm}: Domain \rightarrow 2^{Perms}$ the mapping that grants permissions to protection domains. We \cc{override $\texttt{perm}$ after extending it element-wise and }define access control policy as a mapping $\texttt{policy}: \mathcal{M} \rightarrow 2^{Perms}$, such that $\texttt{policy} = \texttt{perm} \circ \texttt{dom}$.
\end{definition}

%\begin{definition}[\textbf{Check Points}]
By $CheckPoint$ we denote call sites that directly call the method
\textsf{checkPermission}, i.e.,
$CheckPoint = \{\zeta \in CallSite \mid \exists n \in N:  (n, \zeta, n_{check}) \in E \}$. For Example 1, $CheckPoint = \{(checkConnect, 6), (checkAccess, 24)\}$.
%\end{definition}

Let $\phi_{perm}: Perms \rightarrow 2^{AbsCtxt}$ be a mapping from
permissions to the program calling contexts under which permissions
are generated. $Perms$ and $\phi_{perm}$ are generated as follows.
Initially, $Perms = \emptyset$, and $\phi_{perm} = \lambda x.
\emptyset$. For each call site $(m, l) \in CheckPoint$, $l$ contains the expression
``\textit{checkPermission(p)}''. For each $({Type}, loc, c) \in
\texttt{pta}({p})$, the heap allocation site referred to by
$loc$ contains expressions in one of the following
form according to Java API specifications, where \textit{npv} is reference variable of type $T_{perm}$, \textit{target} and \textit{action} are string variables, and ${Type}\in T_{perm}$.
\[
\left\{%
\begin{array}{ll}
    \textit{npv = new~Type(target,action)} & \hbox{(1)} \\
    \textit{npv = new~Type(target)} & \hbox{(2)} \\
        \textit{npv = new~Type()} & \hbox{(3)} \\
\end{array}%
\right.
\]
We augment $Perms$ with a permission $perm$ in the form of 
\begin{itemize}
\item ``${Type}(sv_1,sv_2)$"  if $(sv_1, c_1) \in \texttt{sa}(\textit{target})$, $(sv_2, c_2) \in \texttt{sa}(\textit{action})$, and $c_1 = c_2$ for case (1), and let
$\phi_{perm}(perm)= \phi_{perm}(perm) \cup \{c_1\}$;
\item ``$\textit{Type}(sv)$" if $(sv, c') \in \texttt{sa}(\textit{target})$ for case (2), and let $\phi_{perm}(perm)= \phi_{perm}(perm) \cup \{c'\}$;
\item ``${Type}$" for case (3), and let $\phi_{perm}(perm)= \phi_{perm}(perm) \cup \phi_{method}(m') $ where $m'$ is the method that $loc$ belongs to.
\end{itemize}

\begin{example}
\label{genperm}
Consider $(checkConnect, 6) \in CheckPoint$, we have \\
\begin{tabular}{l}
$~~~~~~~~~~\phi_{meth}(checkConnect) = \{\{\zeta_{(1)}, \zeta_{(3)}\}, \{\zeta_{(2)},\zeta_{(4)}\}\}$ \\
$~~~~~~~~~~\texttt{pta}(p_1) = \{ (SocketPermission, 12, c) \mid c \in \phi_{meth}(checkConnect) \}$\\
\end{tabular} 
\\
Then consider the allocation site at line $12$, we have \\
\begin{tabular}{l}
$~~~~~~~~~~\texttt{sa}(hn) = \{(``jaist.ac.jp/student:8080", \{\zeta_{(2)}, \zeta_{(4)}, \zeta_{(5)}\})\}$ \\$~~~~~~~~~~~~~~~~~~~~\cup \{(``jaist.ac.jp/faculty:8080", \{\zeta_{(1)}, \zeta_{(3)}, \zeta_{(5)}\})\}$\\
$~~~~~~~~~~\texttt{sa}(``connect") = \{(``connect", \{\zeta_{(1)}, \zeta_{(3)}, \zeta_{(5)}\}), (``connect", \{\zeta_{(2)}, \zeta_{(4)}, \zeta_{(5)}\})\}$
\end{tabular}\\
The following permissions can be generated\\
\begin{tabular}{l}
$~~~~~~~~~~``perm_s: SocketPermission(``jaist.ac.jp/student:8080", ``connect")$" \\
$~~~~~~~~~~~\text{with } \phi_{perm}(perm_s) = \{\{\zeta_{(2)}, \zeta_{(4)}, \zeta_{(5)}\}\}$, and \\
$~~~~~~~~~~``perm_f: SocketPermission(``jaist.ac.jp/faculty:8080", ``connect")$"  \\ $~~~~~~~~~~~\text{with } \phi_{perm}(perm_f) = \{\{\zeta_{(1)}, \zeta_{(3)}, \zeta_{(5)}\}\}$\\
\end{tabular}
\\
Consider another check point $(checkAccess, 24)\in CheckPoint$, we can similarly generate permissions \\
\begin{tabular}{l}
$~~~~~~~~~~``perm_{a}: FilePermission(``C:/log.txt", ``write")$  \\ 
$~~~~~~~~~~~\text{with } \phi_{perm}(perm_{a}) = \{\{\zeta_{(1)}, \zeta_{(3)}, \zeta_{(7)}, \zeta_{(8)}\}, \{\zeta_{(2)}, \zeta_{(4)}, \zeta_{(7)}, \zeta_{(8)}\}\}$
\end{tabular}
\end{example}

\section{Problem Formalization}
\label{formalization}

In this section, we fix a context-sensitive call graph $G_{cs} = (G, \phi_{edge})$
where $G = (N, E, s)$, and a dependency graph $G_{dep} = (N_{dep}, E_{dep},$ $ S_{dep})$.

\begin{definition}[\textbf{Valid Call Paths}]
\label{validpath}
\cc{Given a context-sensitive call graph $G_{cs} = (G, \phi_{edge})$
where $G = (N, E, s)$}
We define
\begin{itemize}
\item the set of call paths from $s$ to a node $n \in N$ by
\begin{multline*}
path(n) = \{e_0 e_1\dots e_k \mid \exists k\in \mathbb{N}: n_0 =
s,~n_{k+1} = n, \\
e_i = (n_i, \zeta_i, n_{i+1})\in E  \text{ for each } 0 \leq i \leq
k \}
\end{multline*}
\item the set of call paths from $s$ to $n$ that are truncated by
the node $n_{priv}$ as
\begin{multline*}
tpath(n) = \{ e_0 e_1 e_2 \dots e_k \mid \exists k\in \mathbb{N}: n_0
= s,~n_{k+1} = n, n_0 \rightarrow^* n_{priv},  \\ e_0 = (n_{priv},
\zeta_0, n_1), e_i =(n_i, \zeta_i, n_{i+1}) \text{ for each } 1 \leq
i \leq k\}
\end{multline*}
\item the set of valid call paths from $s$ to a node $n \in N$ by
\begin{multline*}
vpath(n) =  \{ \sigma \in tpath(n) \cup path(n) \mid \exists c \in \phi_{route}(\sigma): c \subseteq sites(\sigma) \text{ and } \\ \sigma \notin vpath(n) \text{ if } \sigma' \in tpath(n) \text{ and } \sigma' \text{ is a suffix of } \sigma\}
\end{multline*}
where $sites(\sigma) = \{ \zeta_0, \dots, \zeta_k \}$ for a call path $\sigma = e_0 e_1\dots e_k $ with $e_i = (n_i,\zeta_i,n_{i+1})$ for each $0 \leq i \leq k$.
\end{itemize}
\end{definition}

\cc{
Consider a call path $\sigma$. Each $c \in \phi_{route}(\sigma)$ is the set of call sites through which $\sigma$ can be valid. Therefore, $\sigma$ is a valid path if there exists $c \in \phi_{route}(\sigma)$ such that $c$ is included in the set of call sites visited by $\sigma$, i.e., $sites(\sigma)$, as defined in Def. \ref{validpath}.}

\begin{example}
Figure \ref{graph} consists of the following valid call paths from $s$ to $n_{check}$: $(1)(3)(6), (2)(4)(6)$, and $((8)(9)(10)$.
\end{example}

\begin{definition}[\textbf{Dependency Paths}]
\cc{Give a dependency graph $G_{dep} = (N_{dep},$ $ E_{dep}, S_{dep})$} We define the set of dependency paths from $S_{dep}$ to
a node $n\in N_{dep}$ by 
\begin{multline*}
dpath(n) = \{e_0 e_1\dots e_k \mid \exists k\in \mathbb{N}: n_0
\in S_{dep},~n_{k+1} = n,\\ e_i = (n_i, n_{i+1})\in E_{dep}  \text{
for each } 0 \leq i \leq k \}
\end{multline*}
and for each dependency path $\pi$, 
\[
extract(\pi) = \left\{%
\begin{array}{ll}
    extract(e) & \hbox{if $\pi = e \in E_{dep}$} \\
    extract(e)extract(\pi') & \hbox{if $\pi = e \pi'$ for some edge $e\in E_{dep}$}
\end{array}%
\right.
\]
where for each edge $e = ((m, l), (m',l')) \in E_{dep}$,
\[
extract(e) = \left\{%
\begin{array}{ll}
    [_{(m,l)} & \hbox{if $(m,l)\in CallSite$ and $e \in E_{inter}$} \\
    ]_{(m',l')} & \hbox{if $(m',l')\in CallSite$ and $e \in E_{inter}$} \\
    \epsilon & \hbox{otherwise}
\end{array}%
\right.
\]
\end{definition}

For a valid call path $\sigma = e_0 e_1\dots e_k $ with $e_i = (n_i,\zeta_i,n_{i+1})$ for each $0 \leq i \leq k$, we define $meths(\sigma) = \{ m_i, m_{i+1} \mid 0 \leq i \leq k\}$. For a dependency path $\pi = e'_0 e'_1 \cdots e'_h$ where $e_i = ((m_i,l_i), (m_{i+1},l_{i+1}))$
for each $0 \leq i \leq h$, we define $meths(\pi) = \{m_i, m_{i+1} \mid 0 \leq i \leq h\}$.

\begin{definition}[\textbf{Relate Valid Call Paths to Permissions}]
\label{relate}
Given a dependency path $\pi = e_0 e_1 \cdots e_k$, where $e_i = ((m_i,l_i), (m_{i+1},l_{i+1}))$
for each $0 \leq i \leq k$. Let $\sigma = e'_0 e'_1 \cdots e'_{h} \in vpath(m_0)$ be a valid call path
from $s$ to $m_0$, where $e'_i =
(m'_i, \zeta_i, m'_{i+1})$ with $\zeta_i = (m'_i,l'_i)$ for each $0 \leq i \leq h$. Let
$\omega_l = [_{(m'_0,l'_0)} \cdots [_{(m'_h,l'_h)}$, and let $w_r = extract(\pi)$.
We say $\pi$ \textbf{matches with} $\sigma$ if $\omega_l \omega_r$ is a well-matched word (every symbol $]_{(m,l)}$ has a matched symbol $[_{(m,l)}$ before it in the word).
The set of all valid call paths that $\pi$ matches with is denoted by $match(\pi)$.

Given a valid call path $\sigma \in vpath(n_{check})$ and a permission $perm \in Perms$. We say $\sigma$ \textbf{relates to} $perm$ if there exists a dependency path $\pi \in dpath(n)$ for some $n\in CheckPoint$, and
\begin{enumerate}[(i)] 
\item there exists a valid call path $\sigma' \in match(\pi)$ such that $meths(\sigma) \subseteq meths(\pi) \cup meths(\sigma')\cup \{n_{check}\}$; and 
\item there exists $c \in \phi_{perm}(perm)$
such that $c \subseteq sites(\sigma')$.
\end{enumerate}
\end{definition}

\cc{Note that both $\pi$ and $\sigma$ in Def. \ref{genpolicy} can be infinitely many.}

In Def. \ref{relate}$, \pi$ matches with $\sigma$ means that $\pi$ and $\sigma$ jointly constitute \textit{a valid inter-procedural data and control flow with respect to the permission} allocated at the initial node of $\pi$. By valid, we mean as usual that methods returns are matched by method calls. Furthermore, if any calling context (i.e., the set of call sites) for allocating a permission is consumed in $\sigma$ matched with $\pi$, we regard that $\pi$ relates to the permission.
We regard a valid call path $\sigma'$ relates to a permission if the set of methods visited by $\sigma'$ is consumed in the valid inter-procedural data and control flow with respect to that permission.

\begin{definition}[\textbf{Policy Generation}]
\label{genpolicy} We define $policy: \mathcal{M} \rightarrow
2^{Perms}$ by, for each valid call path $\sigma \in vpath(n_{check})$, a permission $perm \in Perm$, and $m \in meths(\sigma)$, $perm \in policy(m)$ if $\sigma$ relates to $perm$.
\end{definition}

\begin{example}
\label{ex:formalization}
Consider the valid call path $\sigma: (1)(3)(6) \in vpath(n_{check})$. For the dependency path $\pi: (11)(12)(13)$, $match(\pi) = \{\sigma': (1)(3)(5), \sigma'': (2)(4)(5)\}$ ($\omega_l = [_{\zeta_{(1)}}[_{\zeta_{(3)}}[_{\zeta_{(5)}}$ for $\sigma'$,  $\omega_l = [_{\zeta_{(2)}}[_{\zeta_{(4)}}[_{\zeta_{(5)}}$ for $\sigma'$, and $\omega_r= ]_{\zeta_{(5)}}$ for $\pi$). We have \\
%\begin{center}
\begin{tabular}{l}
$~~~~~meths(\pi) =\{mkSocketPerm, checkConnect\}$\\
$~~~~~meths(\sigma') = \{s, connectFaculty, checkConnect, mkSocketPerm\}$\\
$~~~~~meths(\sigma) = \{s, connectFaculty, checkConnect, n_{check}\}$\\
\end{tabular}
\\
%\end{center}
and therefore $meths(\sigma) \subseteq meths(\sigma') \cup meths(\pi) \cup \{n_{check}\}$. Furthermore, 
by Example \ref{genperm}, we have $\phi_{perm_f} \subseteq sites(\sigma') = \{\zeta_{(1)}, \zeta_{(3)}, \zeta_{(5)}\}$. By Definition \ref{relate} and \ref{genpolicy}, each method of $meths(\sigma)$ holds $perm_f$. In this way we can precisely infer that $\phi_{perm}(connectFactulty)={perm_f}$, $\phi_{perm}(connectStudent)=\{perm_s\}$. 

\cc{Consider the valid call path $\sigma: (9)(11)(13)$. For dependency path $\pi: (17)$, $match(\pi) = \{\sigma': (1)(8)(10)(12), \sigma'': (2)(7)(9)(11)\}$. }
\end{example}

\section{Generating Access Control Policy}
\label{realization}

\begin{definition}[\textbf{Modeling Context-Sensitive Call Graph}]
\label{definition:interprocedural} Given a context-sensitive call graph $G_{cs} = (G, \phi_{edge})$
where $G = (N,E,s)$. We define a conditional pushdown system
$\mathcal{P}_c = (\{\cdot\}, \Gamma, \mathcal{C}, \Delta_c, \{\cdot\}, s)$, where
\begin{itemize}
\item the set of control locations is a singleton $\{\cdot\}$;
\item the stack alphabet $\Gamma \subseteq \mathcal{M} \cup CallSite$;
\item we write $\alpha
\overset{C}{\hookrightarrow}  \omega$ for  $(\cdot, \alpha, C, \cdot, \omega) \in \Delta_c$.
$\Delta_c$ is constructed as follows, for each edge $e=(m, \zeta,
m')\in E$, we have $$m \overset{C(e)}{\hookrightarrow}  m' \zeta \in \Delta_c$$
where
$C(e)=\bigcup_{\{\gamma_0, \cdots, \gamma_k\} \in \phi_{edge}(e)} ~\bigcup_{\{i_0, \cdots, i_{k}\} \in \Pi(\{0,\dots,k\})} \Gamma^* \gamma_{i_0} \Gamma^*  \cdots  \gamma_{i_{k}} \Gamma^*$.
\end{itemize}
\end{definition}

In Def. \ref{definition:interprocedural}, $C(e)$ means that some calling context of the call edge in question is contained in the current call stack.

\begin{definition}[\textbf{Modeling Dependency Graph}]
\label{model_dg}
Give a dependency graph $G_{dep} = (N_{dep}, E_{dep}, S_{dep})$, we define a conditional pushdown system $\mathcal{P}'_c = (\{\cdot\}, \Gamma, \mathcal{C}',$ $ \Delta'_c)$, where \cc{the control location is a singleton $\{\cdot\}$, the stack alphabet $\Gamma \subseteq \mathcal{M} \cup CallSite$, and }$\Delta'_c$ is constructed as follows, for each edge $e = (n, n') \in E_{dep}$ where $n = (m,l)$ and $n' = (m', l')$, we have
\[ m \overset{C(e)}{\hookrightarrow} \epsilon \in \Delta'_c \text{ and } (m',l') \overset{C(e)}{\hookrightarrow} m' \in \Delta'_c\]
where $C(e) = \Gamma^*$ if $(m',l')\in CallSite$.
\end{definition}

\cc{A dependency graph has edges that correspond to statements of method calls. But they have corresponding call edges in the call graph and are considered when modeling context-sensitive call graphs.}

\begin{definition}[\textbf{Program Modeling}]
\label{basic}
We define a conditional pushdown system $\mathcal{P}_{prog} = (\{\cdot\}, \Gamma, \mathcal{C}_{prog}, $ $\Delta_{prog}, \{\cdot\}, s)$ where, $\mathcal{C}_{prog} = \mathcal{C}\cup \mathcal{C}'$, and $\Delta_{prog} = \Delta_c\cup \Delta'_c$, by combining $\mathcal{P}_c$ and $\mathcal{P}'_c$ generated for $G_{cs}$ and $G_{dep}$, respectively.
\end{definition}

\begin{definition}[\textbf{Weight Domain}]
We define a bounded idempontent semiring $\mathcal{S}_{gen} = (D_{gen}, \oplus_{gen}, \otimes_{gen}, \bar{0}, \bar{1})$, where
$D_{gen} \subseteq 2^{2^{\mathcal{M}} \times 2^{\mathcal{M}} \times 2^{\mathcal{M}} \times 2^{CallSite}} \cup \{\bar{0}\}$, and  $\bar{1} = \{(\emptyset, \emptyset, \emptyset, \emptyset)\}$; and
 for any $d, d' \in D_m$, $d \oplus_{gen} d' = d \cup d'$, and
\begin{multline*}
d~ \otimes_{gen} d' = \{ (M_1\cup M'_1 \setminus M'_2, M_2, M_3 \cup M'_3, M_4 \cup M'_4)\mid \\ (M_1, M_2, M_3, M_4) \in d, (M'_1, M'_2, M'_3, M'_4) \in d'\}
\end{multline*}
\end{definition}

One can prove that both $\otimes_m$ and $\oplus_m$ are associative, and $\oplus_m$ is commutative and distributive over $\otimes_m$, which holds for a bounded idempontent semiring.

\begin{definition}[\textbf{Modeling Policy Generation}]
\label{model}
\cc{Given a context-sensitive call graph $G_{cs} = (G, \phi_{edge})$
where $G = (N,E,s, n_{check})$, and a dependency graph $G_{dep} = (N_{dep}, E_{dep}, S_{dep}, F_{dep})$. }We define a conditional weighted pushdown system $\mathcal{W}_{gen} = (\mathcal{P}_{prog}, \mathcal{S}_{gen}, f_{gen})$. For each transition rule $\delta \in \Delta_{gen}$, $f_{gen}(\delta)$ is defined as follows,
\begin{itemize}
\item if $\delta$ is a push rule $m \overset{C}{\hookrightarrow} m' \zeta $,
\[
\left\{
  \begin{array}{ll}
    f_{gen}(\delta) = \{(\{m\}, \Gamma, \emptyset, \{\zeta\})\}, & \hbox{if $m = n_{priv}$;} \\
    f_{gen}(\delta) = \{(\{m\}, \emptyset, \emptyset, \{\zeta\})\}, & \hbox{otherwise}
  \end{array}
\right.
\]
\item if $\delta$ is a pop rule $m \overset{C}{\hookrightarrow} \epsilon$, $f_{gen}(\delta) = \{(\emptyset, \emptyset, \{m\}, \emptyset)\}$.
\item otherwise $f_{gen}(\delta) = \bar{1}$
\end{itemize}
\end{definition}

\begin{definition}[\textbf{Algorithm for Generating Access Control Policy}]
\label{algo_gen}
Given a conditional weighted pushdown system $\mathcal{W}_{gen} = (\mathcal{P}_{prog}, \mathcal{S}_{gen}, f_{gen})$ constructed by Def. \ref{model}. We compute $$result = \texttt{MOVP}(\{\langle \cdot, s \rangle\}, T, \mathcal{W}_{prog})$$ where $T = \{\langle \cdot, n_{check} \omega \rangle \mid \omega \in \Gamma^*\}$.
\begin{enumerate}[(i)]
\item For any $d = (M_1, M_2, M_3, M_4)\in result$, and $perm \in Perms$, we say $perm$ is required by $d$ if there exists $c \in \phi_{perm}(perm)$ such that $c \subseteq M_4$. 
\item For each $m \in M_1\setminus M_3$,
$perm \in policy(m)$ if $perm$ is required by $d$.
\end{enumerate}
\end{definition}

For each $d$ computed in Def. \ref{algo_gen}, $M_4$ is the calling history in terms of call sites of valid inter-procedural data flows constituted by  call paths and dependency paths; $M_1$ contains methods that reside on call paths truncated by $n_{priv}$; $M_2$ is supposed to be $\emptyset$ by our modeling, because $n_{priv}$ can never be the initial node of call graph, and $M_3$ contains finished called methods that do not reside on the current call stack. Therefore $M_1 \setminus M_3$ contains methods residing on the current call stack before a privileged method. The algorithm for generating access control policy precisely corresponds to principles in Def. \ref{relate}. The soundness of our analysis is straightforward given the abstract interpretation in Section \ref{abstraction}.

\begin{example}
\label{ex5}
By Def. \ref{definition:interprocedural}, we have 
$$\Delta_c = \{ \delta_i: m_i  \overset{C(e)}{\hookrightarrow}  m'_i ~ \zeta_{(i)},\mid e: (m_i, \zeta_{(i)}, m'_i), 1 \leq i \leq 10\}$$
By Def. \ref{model_dg}, we have 
$$\Delta'_c=\{\delta: mkSocketPerm  \overset{C}{\hookrightarrow}  \epsilon, \delta': \zeta_{(5)}  \overset{C}{\hookrightarrow} checkConnect\}$$ that is encoded from the edge $(12)$ of dependency graph, where $C = \Gamma^*$. 

We have the following weights for transitions,
\[
\left\{
  \begin{array}{l}
  f_{gen}(\delta_8) = \{(\{n_{priv}\}, \Gamma, \emptyset, \{\zeta_{(8)}\})\}\\
  f_{gen}(\delta) = \{(\emptyset, \emptyset, \{mkSocketPerm\}, \emptyset)\}\\
  f_{gen}(\delta') = \bar{1} \\
  f_{gen}(\delta_i) = \{(\{m_i\}, \emptyset, \emptyset, \{\zeta_{(i)}\})\} \text{ for } i \in \{1,\dots,10\}\setminus \{8\}
  \end{array}%
\right.
\]
By Def. \ref{algo_gen}, we compute $result=\{d_1, d_2, d_3, d_4\}$, where
\[
\left\{
  \begin{array}{l}
d_1 = (M_1, \emptyset, \{mkSocketPerm\}, \{\zeta_{(i)} \mid i \in \{1,3,5,6\}\}) \\
~~~~~~~~~~\text{where } M_1 = \{s, connectFaculty, checkConnect,mkSocketPerm\}\\
d_2 = (M_1, \emptyset, \{mkSocketPerm\}, \{\zeta_{(i)} \mid i \in \{2,4,5,6\}\})\\
~~~~~~~~~~\text{where } M_1 = \{s, connectStudent, checkConnect,mkSocketPerm\}\\
d_3 = (Priv.run, checkAccess\}, \emptyset, \emptyset, \{\zeta_{(i)} \mid i \in \{2,4,7,8,9,10\}\})\\
d_4 = (Priv.run, checkAccess\}, \emptyset, \emptyset, \{\zeta_{(i)} \mid i \in \{1,3,7,8,9,10\}\}) \\
\end{array}%
\right.
\]
%\end{center}
We have that $perm_f$ is required by $d_1$,  $perm_s$ is required by $d_2$, and $perm_{a}$ is required by $d_3$ and $d_4$. We can further infer permissions possessed by each method by (ii) of Def. \ref{algo_gen}.
\end{example}

\section{Discussions}
\label{discussion}

\subsection{Context-Sensitive Points-to and String Analysis}
Context-sensitive points-to and string analysis play a crucial role in our analysis framework. It is not hard to adapt the existing analysis to fit our needs.

It is direct to adapt static analysis by WPDSs to our setting, because WPDSs have the advantage of handling data flow queries as regular languages of pushdown configurations, and regular stack configurations naturally encode our abstraction of calling contexts. For instance, for each reference variable $v$ of the method $m$, we can compute $\textsf{pta}(v)$ = $\underset{T_{ctxt}}{\bigcup}\textsf{MOVP}(S,T_{ctxt})$ where $S$ is the source configurations, and $T_{ctxt} = \{\langle v, m \omega \rangle \mid \Sigma(\omega) \subseteq ctxt\}$ for each $ctxt \in \phi_{method}(m)$. \cc{For string-analysis, the analysis in \cite{Choi2006} based on context-cloning ($k$-CFA) can be reformulated in the framework of WPDSs, and then adapted to a context-sensitive analysis similarly to points-to analysis.}

To adapt cloning-based analysis to our needs, we can turn to the following approach that is line with context-cloning: 
given a call graph $G=(N,E,s)$ (or  graphs which product with call graph as the starting point of the analysis). We construct another graph $G_{clone}=(N_{clone}, E_{clone})$, where $N_{clone} \subseteq 2^{AbsCtxt} \times N$ is the set of nodes,  $E_{clone} \subseteq N_{clone} \times CallSite \times N_{clone}$ is the set of edges, and we have
$(i)$ $(c, n) \in N_{clone}$ for $c \in \phi_{method}(n)$, and $n\in N$; and $(ii)$ $((c, n), (c', n')) \in E_{clone}$ if $c \subseteq c'$ and $(n, n')\in E$ for $(c, n), (c', n') \in N_{clone}$.

One obtains context-sensitive analysis by applying context-insensitive analysis to $G_{clone}$, e.g., the points-to analysor Spark \cite{Lhotak2003} can be adapted by cloning its points-to graph in this manner, and string analysor JSA (Java String Analyzer) \cite{Christensen2003} can be lifted to a context-sensitive analysis of our framework by cloning its front-end flow graph, with no need to modify the back-end analysis engine.

\subsection{Policy Checking}
Another popular need in access rights analysis is checking whether the program function properly given an access control policy, e.g., codes from trusted domains always pass access control or may fail. The answer can either help detect redundant inspection points or refine the given policies.

One approach to policy checking is first generating an access control policy as described before, and then check whether the given policy consumes the required policy.  Formally,
given a $policy: \mathcal{M} \rightarrow 2^{Perms}$ and a
$policy': \mathcal{M} \rightarrow 2^{Perms}$ generated by
Def.\ref{genpolicy}. Stack inspection triggered in the
program always succeed if $policy'(m) \subseteq policy(m)$ for
each $m \in \mathcal{M}$, and may fail otherwise.

However, one problem for this approach is that, we may reject a benign security policy if it is more precise than the generated one. After all, the access control policy automatically generated is an over-approximation of the \text{minimal} policy by sound static analysis. 

Instead of generating the minimal policy in advance, an alternative is to check on-demand at stack inspection points whether all methods in the current call stack are granted required permissions. The two approaches to policy checking are in line with the two ways of implementing stack inspection mechanism by virtual machines in an either \textit{eager} or \textit{lazy} manner.

\section{Related Work}
\label{related}

From the theoretical aspect, Banerjee et al. in \cite{Banerjee2001} gived a denotational semantics and hereby proved the equivalence of eager and lazy evaluation for stack inspection. They further proposed a static analysis of safety property, and also identified program transformations that help remove redundant runtime access control checks. The problem to decide whether a program satisfies a given policy properties via stack inspection, was proved intractable in general by Nitta et al. in \cite{Nitta2001}. They showed that there exists a solvable subclass of programs which precisely model programs containing \textsf{checkPermission} of Java 2 platform. Moreover, the study concluded the computational complexity of the problem for the subclass is linear time in the size of the given program.

Chang et al. \cite{Chang2006} provided a backward static analysis to approximate redundant permission checks with must-fail stack inspection and success permission checks with must-pass stack inspection. This approach was later employed in a visualization tool of permission checks in Java \cite{Kim2007}.  But the tool didn't provide any means to relieve users from the burden of deciding access rights.  In addition to a policy file, users were also required to explicitly specify which methods and permissions to check.
Two control flow forward analysis, Denied Permission Analysis and Granted Permission Analysis, were defined by Bartoletti et al. \cite{Bartoletti2001} \cite{Bartoletti2004} to approximate the set of permissions denied or granted to a given Java bytecode at runtime. Outcome of the analysis were then used to eliminate redundant permission checks and relocate others to more proper places in the code.

Koved et al. in \cite{Koved2002} proposed a  context-sensitive, flow-sensitive, and context-sensitive (1-CFA) data flow analysis to automatically estimate the set of access rights required at each program point. In spite of notable experimental results, the study suffered from a practical matter, as it does not  properly handle strings in the analysis.
Being a module of privilege assertion in a popular tool -- IBM Security Workbench Development for Java (SWORD4J) \cite{Habeck2008}, the interprocedural analysis for privileged code placement \cite{Pistoia2005} tackled three neat problems: identifying portions of codes that necessary to make privileged, detecting tainted variables in privileged codes, and exposing useless privileged blocks of codes, by utilizing the technique in \cite{Koved2002}.

In aforementioned works, they all assume permissions required at every  \textsf{checkPermission(perm)} point. That is, they either ignored or employed limited computation of \textsf{String} parameters. Correspondingly, the access rights analysis become too conservative, e.g., many false alarms may be produced in policy checking.

To the best of our knowledge, the modular permission analysis proposed in \cite{Geay2009} is the most relevant to our work . On one hand, it was also concerned with automatically generating security polices for any given program, with particular attention on the principle of least privilege. On the other hand,  they were the first to attempt to reflect  the effects of string analysis in access rights analysis in terms of slicing. A modular analysis algorithm is proposed to achieve the practical scalability, and the authors developed a tool Automated Authorization Analysis (A3)  to assess the precision of permission requirements for stack inspection. However, their algorithms are based on a context-insensitive call graph and the analysis results can be polluted by invalid call paths. Moreover, their slicing algorithms are also context-insensitive.

\section{Conclusions}
\label{conclusion}

We have presented a systematic approach to automated generation of access control policies for a given Java program. The techniques are abstract interpretation based context-sensitive static program analysis. We define an abstract interpretation on program calling contexts, and all analysis modules required in access rights analysis are hereby glued together in an unified analysis framework. Given such an abstract interpretation, we generate permissions that are possibly involved in access rights analysis. In our analysis, the program is modelled by combining a context-sensitive call graph with a dependency graph of the target program and we are therefore able to precisely identify permission requirements at checkpoints of stack inspection. We expect a good precision of our analysis due to its context-sensitive nature. 
A public tool that can automatically generating security policies for Java applications doesn't exist so far. It would be interesting to put the analysis techniques proposed in the paper into practice by settling scalability. As the first step, we are implementing an efficient model checking algorithm for Conditional WPDSs, tailored for algorithms presented in the paper. Although stack inspection is widely adopted as a simple and practical model in stack-based access control, it has a number of inherent flaws, e.g., an unauthorized code which is no longer in the call stack may be allowed to affect the execution of security-sensitive code.
A worth highlighting alternate model is IBAC (Information-based Access Control) proposed by Pistoia et al. in \cite{Pistoia2007} for programs. It would be interesting to extend the analysis framework to analyse IBAC security policy.

\bibliographystyle{abbrv}
\bibliography{newref}

\begin{thebibliography}{10}

\bibitem{Banerjee2001}
A.~Banerjee and D.~A. Naumann.
\newblock {A simple semantics and static analysis for Java security}.
\newblock Technical report, Stevens Institute of Technology, 2001.

\bibitem{Bartoletti2001}
M.~Bartoletti and P.~Degano.
\newblock {Static analysis for stack inspection}.
\newblock {\em Electronic Notes in Theoretical Computer Science}, 54:706--80,
  Aug. 2001.

\bibitem{Bartoletti2004}
M.~Bartoletti and P.~Degano.
\newblock {Stack inspection and secure program transformations}.
\newblock {\em International Journal of Information}, 2004.

\bibitem{Chang2006}
B.~Chang.
\newblock {Static check analysis for Java stack inspection}.
\newblock {\em ACM SIGPLAN Notices}, 41(3):40, Mar. 2006.

\bibitem{Christensen2003}
A.~S. Christensen, A.~M{\o}ller, and M.~I. Schwartzbach.
\newblock Precise analysis of string expressions.
\newblock In {\em Proceedings of the 10th international conference on Static
  analysis}, SAS'03, pages 1--18, Berlin, Heidelberg, 2003. Springer-Verlag.

\bibitem{Geay2009}
E.~Geay, M.~Pistoia, B.~G. Ryder, and J.~Dolby.
\newblock {Modular string-sensitive permission analysis with demand-driven
  precision}.
\newblock {\em 2009 IEEE 31st International Conference on Software
  Engineering}, pages 177--187, 2009.

\bibitem{Habeck2008}
T.~Habeck, L.~Koved, M.~Pistoia, and Y.~Heights.
\newblock {SWORD4J : Security WORkbench Development environment 4 Java}.
\newblock Technical report, IBM, 2008.

\bibitem{Kim2007}
Y.~Kim.
\newblock {Visualization of permission checks in Java using static analysis}.
\newblock {\em Information Security Applications}, pages 133--146, 2007.

\bibitem{Koved2002}
L.~Koved, M.~Pistoia, and A.~Kershenbaum.
\newblock {Access rights analysis for Java}.
\newblock In {\em Proceedings of the 17th ACM SIGPLAN conference on
  Object-oriented programming, systems, languages, and applications},
  volume~37, pages 359----372. ACM, Nov. 2002.

\bibitem{Lhotak2003}
O.~Lhot\'{a}k and L.~Hendren.
\newblock Scaling {J}ava points-to analysis using spark.
\newblock In {\em Proceedings of the 12th international conference on Compiler
  construction}, CC'03, pages 153--169, Berlin, Heidelberg, 2003.
  Springer-Verlag.

\bibitem{Li2010}
X.~Li and M.~Ogawa.
\newblock {Conditional weighted pushdown systems and applications}.
\newblock {\em Proceedings of the ACM SIGPLAN 2010 workshop on Partial
  evaluation and program manipulation - PEPM '10}, page 141, 2010.

\bibitem{Nitta2001}
N.~Nitta and Y.~Takata.
\newblock {An efficient security verification method for programs with stack
  inspection}.
\newblock {\em Computer and Communications Security}, pages 68--77, 2001.

\bibitem{Pistoia2007}
M.~Pistoia, A.~Banerjee, and D.~Naumann.
\newblock {Beyond stack inspection: A unified access-control and
  information-flow security model}.
\newblock {\em Security and Privacy, 2007}, 2007.

\bibitem{Pistoia2005}
M.~Pistoia, R.~Flynn, and L.~Koved.
\newblock {Interprocedural analysis for privileged code placement and tainted
  variable detection}.
\newblock {\em ECOOP 2005-Object-Oriented}, pages 362--386, 2005.

\bibitem{Reps2005}
T.~Reps, S.~Schwoon, S.~Jha, and D.~Melski.
\newblock {Weighted pushdown systems and their application to interprocedural
  dataflow analysis}.
\newblock {\em Science of Computer Programming}, 58(1-2):206--263, Oct. 2005.

\bibitem{Schwoon2002}
S.~Schwoon.
\newblock {\em {Model-Checking Pushdown Systems}}.
\newblock PhD thesis, Technische Universitat Munchen, 2002.

\end{thebibliography}

\end{document}